\documentclass[num-refs]{wiley-article}

\usepackage{siunitx}
\usepackage{graphicx}
\usepackage{caption}
\usepackage{subcaption}
\usepackage{multirow}
\usepackage{xcolor}
\usepackage{anyfontsize}

\usepackage{todonotes}

\papertype{Original Article}
\paperfield{Journal Section}

\title{Generalization in birdsong classification: \\ impact of transfer learning methods and \\ dataset characteristics}

\author[1]{Burooj Ghani}
\author[1]{Vincent J. Kalkman}
\author[2]{Bob Planqué}
\author[2]{Willem-Pier Vellinga}
\author[3]{Lisa Gill}
\author[1, 4]{Dan Stowell}

\affil[1]{Naturalis Biodiversity Center, Leiden, The Netherlands}

\affil[2]{Xeno-Canto Foundation, The Netherlands}
\affil[3]{Landesbund fur Vogel- und Naturschutz, Hilpoltstein, Germany}
\affil[4]{Department of Cognitive Science and Artificial Intelligence, Tilburg University, Tilburg, The Netherlands}

\corraddress{Burooj Ghani, Naturalis Biodiversity Center, Leiden, Netherlands}
\corremail{burooj.ghani@naturalis.nl}

\fundinginfo{Burooj Ghani is supported by the EU-funded projects Mambo and Guarden. Specifically, Mambo is funded under grant agreement No. 101060639, and Guarden is supported by grant agreement No. 101060693.}

\runningauthor{Ghani et al.}

\begin{document}

\begin{frontmatter}
\maketitle

\begin{abstract}
Animal sounds can be recognised automatically by machine learning, and this has an important role to play in biodiversity monitoring. Yet despite increasingly impressive capabilities, bioacoustic species classifiers still exhibit imbalanced performance across species and habitats, especially in complex soundscapes. In this study, we explore the effectiveness of transfer learning in large-scale bird sound classification across various conditions, including single- and multi-label scenarios, and across different model architectures such as CNNs and Transformers. Our experiments demonstrate that both fine-tuning and knowledge distillation yield strong performance, with cross-distillation proving particularly effective in improving in-domain performance on Xeno-canto data. However, when generalizing to soundscapes, shallow fine-tuning exhibits superior performance compared to knowledge distillation, highlighting its robustness and constrained nature. Our study
further investigates how to use multi-species labels, in cases where these are
present but incomplete. We advocate for more comprehensive labeling practices within the animal sound community, including annotating background species and providing temporal details, to enhance the training of robust bird sound classifiers. These findings provide insights into the optimal reuse of pretrained models for advancing automatic bioacoustic recognition.

\end{abstract}
\end{frontmatter}

\section{Introduction}
Biodiversity is declining globally at an alarming rate, making it crucial for research to provide reliable estimates of the health of global ecosystems~\cite{raven2020maintaining}. While several modalities of data streams are in place to cater to this formidable challenge of monitoring biodiversity, sound-based monitoring methods coupled with machine learning provide a robust, fast, and scalable solution~\cite{rutz2023using, rasmussen2024sound}. Population monitoring depends on estimating species occupancy in specific locations to assess changes over time. Animal vocalizations serve as a good (though incomplete) non-invasive proxy for occupancy, as their presence in sound clips indicate species presence~\cite{wood2019detecting, rasmussen2024sound}. The advent of deep learning in bioacoustics has sparked the development of advanced machine learning methods, enabling automated analysis of large acoustic datasets~\cite{stowell2022computational, bick2024national}. Using these methods, researchers can automatically detect and categorize animal vocalizations, significantly saving time and effort while facilitating the study of lesser-known species~\cite{brunk2023quail}.

Birdsong classification has gained significant attention due to its applications in biodiversity monitoring and conservation, enabled by the increasing accuracy of birdsong species classifiers. 
\citeauthor{kahl2021birdnet}~\cite{kahl2021birdnet} introduced BirdNET, a neural network model specifically designed for bird sound classification. Trained on a large dataset of bird sounds, BirdNET v2.2 can recognize over 3000 bird species.~\citeauthor{stowell2019automatic}~\cite{stowell2019automatic} reported the application of modern machine learning techniques to the task of acoustic bird detection, demonstrating that very high retrieval rates can be achieved. Their work highlighted the potential of deep learning in remote monitoring, overcoming the challenges posed by varying environments and species diversity.

However, the development of these tools typically depends on two key resources.
%
The first is the availability of well-annotated training data, since the strong performance of deep learning typically requires much larger data volumes than prior methods. There is ample training data available for some taxonomic groups: for bird species, these are overwhelmingly sourced from community collections like Xeno-canto\footnote[2]{\url{https://xeno-canto.org}} or the Macaulay Library\footnote[3]{\url{https://www.macaulaylibrary.org}}. These collections, however, suffer from some limitations. Firstly, sufficient training data is often lacking for rare and endangered species, which are often the prime target of conservation efforts~\cite{stowell2019automatic}.
Secondly, the labelling (annotation) of collections may also be incomplete or approximate, lacking precise temporal details and omission of species audible in the background.
Despite these limitations, such community audio collections are the key ingredient in all high-performing general-purpose birdsong classifiers.

The second key resource, increasing in importance, is the availability of
machine learning models trained by others.
\textit{Transfer learning}---re-using the information encoded in a previously-trained machine learning algorithm---has been widely adopted in the field of audio classification due to its effectiveness in addressing the limitations of small and imbalanced datasets~\cite{ghani2023global}. Transfer learning is also increasingly crucial as the idea of \textit{foundation models} rapidly evolves, offering large-scale, flexible pre-training adaptable to numerous downstream tasks. Transfer learning enables these foundation models, while scale amplifies their effectiveness~\cite{bommasani2021opportunities}. In bioacoustics, these models can learn broad, generalizable representations, useful for tasks like species detection, classification, and individual or call type identification, even extending to unknown taxa~\cite{van2024birds}.

The dominant approach to transfer learning is \textit{fine-tuning}, which means additional iterations of training applied to the pre-trained neural network using a new dataset.
 Notably, models pre-trained on large audio datasets have shown significant improvements in performance when fine-tuned for specific tasks. 
 
Researchers have successfully applied transfer learning to various audio tasks, such as speech recognition~\cite{wang2015transfer}, music genre classification~\cite{ghosal2018music}, and environmental sound event classification~\cite{tsalera2021comparison}. \citeauthor{gemmeke2017audio}~\cite{gemmeke2017audio} introduced AudioSet, a large-scale dataset containing over 2 million human-annotated 10s sound clips drawn from YouTube videos. This dataset has served as a benchmark for training robust audio classification models. For instance, \citeauthor{hershey2017cnn}~\cite{hershey2017cnn} explored various convolutional neural network (CNN) architectures pre-trained on AudioSet and demonstrated their efficacy in various audio recognition tasks. \citeauthor{gong2021psla}~\cite{gong2021psla} proposed the PSLA model, a CNN-based approach trained on AudioSet. PSLA's architecture includes specialized layers for capturing both local and global features of audio signals, resulting in superior performance on various sound event classification benchmarks. \citeauthor{koutini2021efficient}~\cite{koutini2021efficient} presented the PaSST model, which leverages an optimized transformer-based architecture for audio classification. Pre-trained on AudioSet, PaSST effectively captures long-term dependencies in audio signals, making it suitable for complex audio classification tasks. 

Transfer learning has also been utilized in bioacoustics. Çoban, et al. employ the VGGish model, pre-trained on AudioSet, to classify ecoacoustic recordings from the North Slope of Alaska into eight different sound classes, including birdsongs and waterfowl sounds~\cite{ccoban2020transfer}. Ghani, et al. showed in~\cite{ghani2023global} that embeddings derived from global bird sound classifiers can be effectively used for few-shot transfer learning across various taxonomic groups, including marine mammals, bats, and frogs. Lauha, et al. demonstrated that finetuning a bird sound recognition model on local data improved the model performance in the specific acoustic enviroment~\cite{lauha2022domain}. In~\cite{dufourq2022passive} Dufourq, Emmanuel, et al. demonstrated that transfer learning, specifically by adapting CNN architectures pre-trained on ImageNet, can effectively simplify the creation of bioacoustic classifiers for passive acoustic monitoring (PAM). The study highlighted that this approach reduces the complexity of CNN design and hyper-parameter tuning, making it more accessible and efficient even with very small datasets.  

As an alternative to fine-tuning, the information could be transferred from the pretrained model to an entirely new model.
\textit{Knowledge distillation} is a technique where a new ``student'' model is trained to replicate the behavior of an existing ``teacher'' model~\cite{gou2021knowledge, hinton2015distilling}. This approach has been employed to create small and efficient models that retain high performance while being more suitable for deployment in resource-constrained environments~\cite{priebe2024efficient}.
This method has been widely adopted in various domains, including image and audio classification, to compress large models without significant loss in accuracy. Knowledge distillation has shown promising results in audio classification tasks. For instance, Choi et al.~\cite{choi2022temporal} applied knowledge distillation to develop a compact audio event classification model that achieved competitive performance with reduced computational requirements. 

Our goal is to develop a highly accurate classifier for European bird species.
While training from multiple diverse data sources is recommended for strong generalisation, training from a single common dataset is important for research study, benchmarking, and for pretraining.
Xeno-canto is a uniquely valuable birdsong collection and is used as the core data to train all successful general-purpose birdsong classifiers. Its role in the development of computational bioacoustics has been analogous to the role of ImageNet in computer vision.
In the present study we investigate how to train a model on Xeno-canto alone, the best to generalise to soundscape data.

To train a sound recognition algorithm, the key resource is a large annotated dataset. 
 However other datasets and models trained for related tasks are available and these could serve as a starting point allowing for building better models in a more efficient way. For a given machine learning goal, the question is thus how best to (re-)use these resources, since multiple strategies are available even within the domain of transfer learning. In this paper we explore this by empirically comparing strategies of fine-tuning against knowledge distillation.
The outcome is a high-performing European bird species classifier, trained using fully open Xeno-canto data.

\section{Method}

\subsection{Transfer learning}

Transfer learning is a machine learning technique where knowledge gained from one domain is applied to another domain.
``Knowledge'' here is a shorthand term to express that a trained system has stored latent information about how to extract general or specific information from its input data, based on the training data's characteristic patterns and distributions.
The approach leverages a model developed for one task as a starting point for another, related task, often compensating for limited or inadequate data in the destination domain~\cite{tsalera2021comparison}. The primary idea is to utilize the patterns learned from a large dataset in the source domain and fine-tune them to improve performance on a different but related problem in the target domain.
In the context of deep learning, transfer learning most commonly involves using a pre-trained network (trained on a large dataset such as ImageNet or AudioSet) and fine-tuning it on a smaller, task-specific dataset.
Note however that there can be many ways to achieve transfer learning. For example, Ntalampiras (2018) extracted features from a (non-deep) classifier trained on music, as input to a bird classifier~\cite{ntalampiras2018bird}.
 
\begin{figure}[ht!] 
\centering
\includegraphics[width=4in]{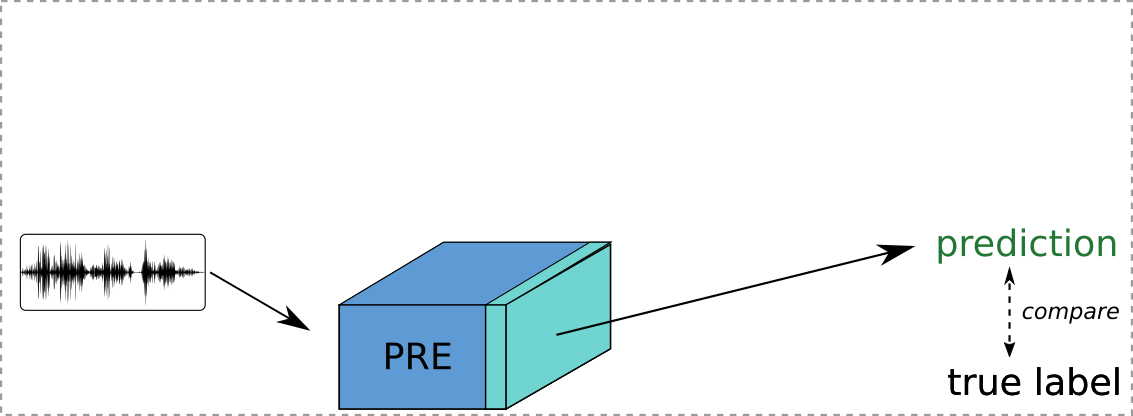}
\includegraphics[width=4in]{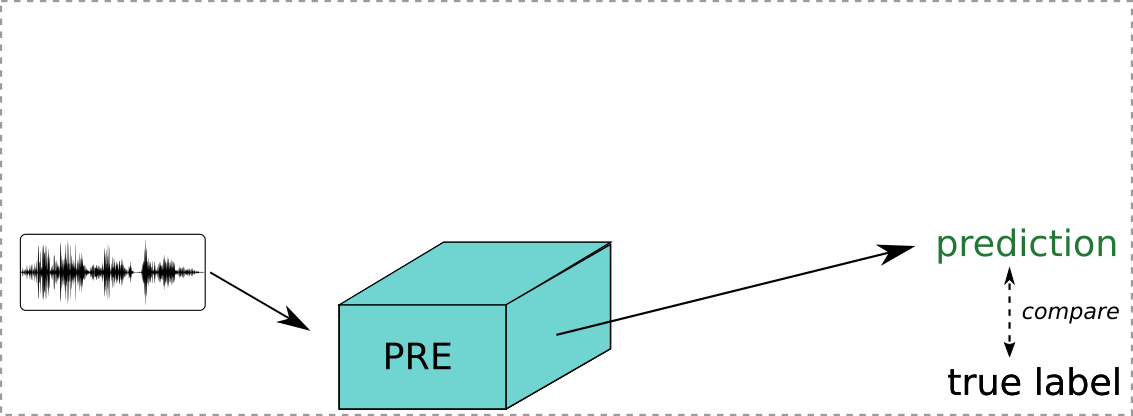}
\includegraphics[width=4in]{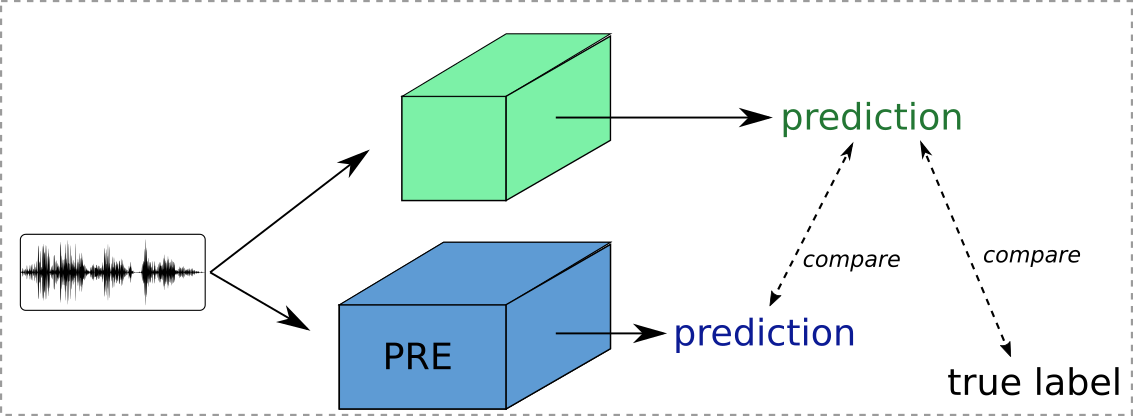}
\caption{Transfer learning strategies. Light-coloured blocks are neural networks being trained; dark-coloured blocks are `frozen' and unchanging during transfer learning. \textbf{Shallow fine-tuning (top)} uses most of the pretrained model as a fixed feature extractor, retraining the final layer(s) on the new dataset. \textbf{Deep fine-tuning (middle)} retrains all layers. \textbf{Knowledge distillation (lower)} differs from both of these, training a new model to produce the same outputs as the teacher model, and also to match ground-truth labels when they are available.}
\label{fig:transfermethods}
\end{figure}

Birdsong classification is a challenging task due to the limited availability of labeled bird sound recordings and the complexity and variability of bird vocalizations. Transfer learning is particularly useful in this context for several reasons. First, data scarcity is a common issue, as birdsong datasets are often small and imbalanced. Transfer learning allows us to leverage pre-trained models that have been trained on large datasets, mitigating this problem. Second, these pre-trained models have learned to extract rich features that capture important patterns and characteristics in audio signals, which can significantly enhance the accuracy of birdsong classification. Additionally, starting with a pre-trained model reduces the time and computational resources needed to train a model from scratch, as the initial layers already contain useful representations. In this study, we explore different transfer learning strategies, including deep fine-tuning, shallow fine-tuning, and knowledge distillation, to capitalize on the advantages of using pre-trained models for birdsong classification. These strategies not only address data scarcity and improve feature richness but also streamline the training process, making transfer learning particularly beneficial for this task.








We next describe the transfer learning strategies considered in the present work.

\subsubsection{Deep finetuning}

Deep fine-tuning involves adjusting all layers of a pre-trained model on a specific dataset, in this case, the birdsong dataset. This comprehensive adjustment enables the model to better capture and adapt to the unique characteristics of bird sounds, adapting its ``early'' and ``late'' processing stages and thus potentially improving its performance on this specialized task. However, this approach has two potential drawbacks. Retraining the entire network may be computationally costly, especially as larger and larger networks are introduced. The method also risks overfitting to the new dataset, since there are many free parameters: if all layers are free to change, the pretrained knowledge may be fully erased by the later training, a phenomenon referred to as ``catastrophic forgetting''~\cite{mccloskey1989catastrophic, goodfellow2013empirical}. To mitigate this, deep fine-tuning needs careful implementation and/or a relatively large dataset.

The optimisation for deep fine-tuning is the same as for ordinary training of a deep neural network, in which we define a loss function $L$ dependent on the input (spectrogram) $X$ and the ground-truth label $y$:
\begin{equation*}
\text{L(X,y)} = \text{L}_{g}(\delta(f(X)), y)
\end{equation*}
Here $L_g$ is a standard loss function such as cross-entropy, and $f$ representing the neural network. The final nonlinearity of the neural network is shown explicitly as the function $\delta$, which we will later refer to, separately from the unnormalised outputs (``logits'') of $f$. The parameters of $f$ (not shown) are chosen to minimise $L$.

In the context of deep learning, fine-tuning pre-trained models has been shown to yield significant improvements in various tasks. For instance,~\citeauthor{howard2018universal}~\citep{howard2018universal} demonstrated that fine-tuning the entire model, as opposed to only the final layers, can result in substantial gains in performance on text classification tasks . Similarly, in the field of computer vision, full fine-tuning of convolutional neural networks (CNNs) has been proven effective in adapting models to new domains, as evidenced by~\citeauthor{yosinski2014transferable}~\citep{yosinski2014transferable}, who highlighted the benefits of fine-tuning across different tasks and datasets .

Moreover, recent advancements in fine-tuning techniques, such as those described by~\citeauthor{raffel2020exploring}~\citep{raffel2020exploring} in their work on the T5 model, underscore the importance of appropriately scaling the model and dataset size to achieve optimal results. Applying these principles to the task of birdsong recognition, deep fine-tuning has the potential to harness the power of pre-trained models to enhance performance, provided that sufficient data and computational resources are available to support the process.

\subsubsection{Shallow finetuning}

In shallow fine-tuning, only the final layers of a pre-trained model are adjusted to fit on the target dataset, while the early layers remain unchanged.
The unchanged part of the pre-trained model can be considered as a feature extractor, providing a rich set of learned features that should be useful for the new task.
This method is less computationally intensive compared to deep fine-tuning, since only a small/shallow part of the network needs to be trained.
It can be particularly effective when working with smaller datasets, as it is strongly constrained to leverage the general features previously learned by the lower layers of the model.

To express shallow fine-tuning mathematically, we separate the network $f$ into two parts:
\begin{equation*}
\text{L(X,y)} = \text{L}_{g}(\delta(f_2(f_1(X)), y)
\end{equation*}
where $f_1$ is constant, and the parameters of $f_2$ are optimised.

Shallow fine-tuning has been demonstrated to be effective in various contexts.~\citeauthor{white2023one}~\cite{white2023one} showed that adapting trained CNNs to new marine acoustic environments by only fine-tuning the top layers allowed the models to effectively generalize to different conditions without extensive retraining . This strategy not only saves computational resources but also reduces the risk of overfitting, making it suitable for tasks with limited data.

Additionally,~\citeauthor{ghani2023global}~\cite{ghani2023global} highlighted the benefits of using global birdsong embeddings to enhance transfer learning for bioacoustic classification. By leveraging pre-trained embeddings and fine-tuning only the top layers, their approach achieved superior performance with minimal computational overhead. Similarly, ~\citeauthor{williams2024leveraging}~\cite{williams2024leveraging} emphasized the advantages of using pre-trained models on diverse sound datasets, including tropical reef and bird sounds, to improve transfer learning in marine bioacoustics. Their work supports the efficacy of shallow fine-tuning in adapting models to new acoustic environments .

These studies underscore that shallow fine-tuning can effectively utilize the robust, general features captured by the lower layers of pre-trained models, making it a practical and efficient approach for tasks like birdsong recognition, especially when computational resources and dataset sizes are limited.

\subsubsection{Knowledge distillation}
Knowledge distillation is a technique where a pre-trained model is not itself changed: instead, it acts as a ``teacher'' and guides the training of a student model (Figure \ref{fig:transfermethods}, lower panel). In doing so the knowledge is transferred to the student model, to obtain a competitive performance versus the teacher model~\cite{gou2021knowledge, hinton2015distilling}.
This is often applied by training the student on ``soft targets'' from the teacher model: the fuzzy predictions rather than binarised 1-or-0 decisions. We interpret this as training the student to approximate the continuously-varying prediction function that the teacher outputs.

While model compression for efficiency and speed is a significant motivation behind this process, i.e.\ training smaller models from large~\cite{priebe2024efficient}, improved generalization and transfer learning are also key benefits. Learning from the nuanced information in the teacher's soft targets helps the student model to learn the rich similarity information between classes that the teacher model has established~\cite{schmid2023efficient, hinton2015distilling}. Interestingly, recent research has shown that cross-model distillation between different architectures, such as convolutional neural networks (CNNs) and transformer models, can yield superior performance, leveraging the strengths of both types of models to create a more robust and effective student model~\cite{gong2022cmkd}.
This can be interpreted as one model learning to approximate information encoded not only in the data, but also in the biases implicit in the architecture of a very different network.

The original knowledge distillation framework was proposed by~\cite{hinton2015distilling}.
Although diverse and more complex attention distillation techniques have since been studied~\cite{touvron2021training}, 
the authors in~\cite{gong2022cmkd} reported that the original method is more effective, provided that certain details such as ``consistent teacher'' training are used \cite{beyer2022knowledge}.
Therefore, we employ this approach. During the training of the student model, we use the same input with ``consistent teaching'', meaning that the teacher and student are exposed to identical mel spectrograms, even when data augmentation is used to modify the training data.
To train the student model the following loss is minimized, composed of two parts~\cite{gong2022cmkd}:


\begin{equation*}
\text{L(X,y)} = \lambda \text{L}_{g}(\delta(f_s(X)), y) + (1 - \lambda) \text{L}_{kd}(\delta(f_s(X)), \delta(f_t(X) / \tau))
\end{equation*}
The coefficient \(\lambda\) represents the weighting whose value, following~\cite{gong2022cmkd}, we fix at $0.5$. The terms \(\text{L}_g\) and \(\text{L}_{kd}\) correspond to the ground truth and distillation losses, respectively. In other words, the knowledge distillation can be seen as a weighted sum of ground truth loss \(\text{L}_g\) and distillation loss \(\text{L}_{kd}\). The symbol \(\delta\) denotes the activation function, while \(f_s\) and \(f_t\) are the student and teacher models, respectively. The parameter \(\tau\) stands for the "temperature". It is important to note that the teacher model remains frozen during the training of the student model.

We employ the cross-entropy (CE) loss as \(\text{L}_g\), combined with the softmax activation function, for multi-class classification (single-label AvesEcho v0). For the multi-label classification (AvesEcho v1), we utilize the binary cross-entropy (BCE) loss as \(\text{L}_g\) with the sigmoid activation function. Additionally, we use Kullback–Leibler divergence as the knowledge distillation loss \(\text{L}_{kd}\) for multi-class classification, and binary cross-entropy (BCE) loss for multi-label classification.

\subsection{Audio data}
To build a model allowing for the monitoring of European breeding birds we selected species in the 2nd European Breeding Bird Atlas (EBBA2)~\cite{keller2020european} identified as breeding in Europe. This also includes all birds normally wintering or migrating in our target geography (running up to the Ural in the east). This list includes all birds found breeding in Europe in the period 2013-2017 including established non-native species.
\begin{figure}[ht!] 
\centering
\includegraphics[width=5.5in]{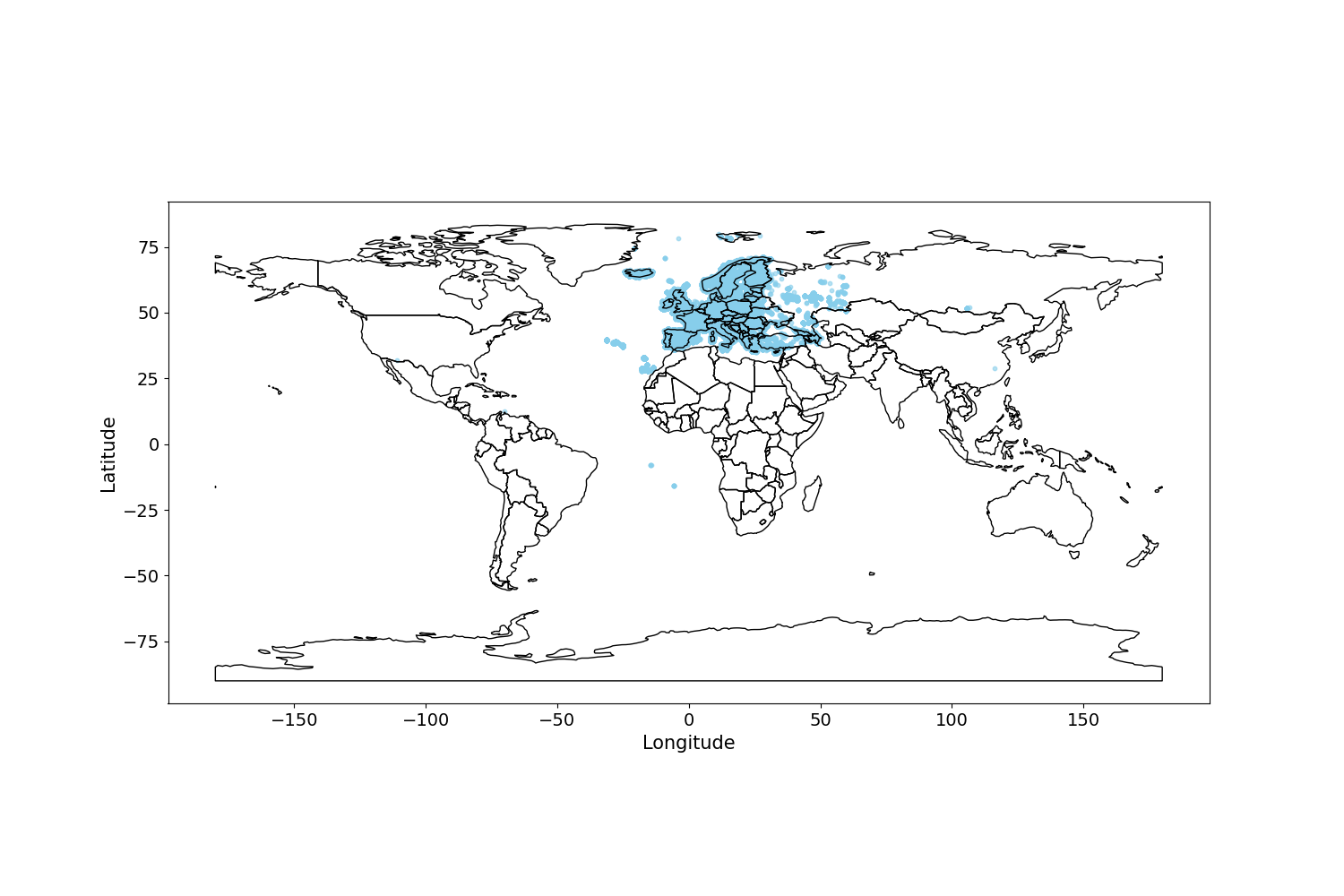}
\caption{Geographic distribution of our data sourced from Xeno-canto.}
\label{distribution_xc_geo}
\end{figure}

We then downloaded corresponding audio data from Xeno-canto~\cite{vellinga2015xeno}, the citizen-science open data source for animal sound recordings. We used the Xeno-canto API to select sound recordings preferentially of high quality, and recorded in Europe. Xeno-canto (XC) is the largest open access database of nature sounds and available recordings cover all European breeding birds, although for some very rare species the number of recordings is limited. Xeno-canto has the benefit that the overall quality of recordings is good and the identification is reliable due to peer-review by other workers. A further advantage is that the data is largely open access so that future models can make use of the same source of training data.
\begin{table}[h!]
\centering
\rowcolors{0}{}{} 
\begin{tabular}{>{\raggedright\arraybackslash}p{2.5cm} > {\raggedright\arraybackslash}p{2cm} >{\raggedright\arraybackslash}p{2cm} >{\raggedright\arraybackslash}p{2cm} c}
\hline
\textbf{Dataset source}  & \textbf{Multi-label} & \textbf{Classes} & \textbf{Sound Files} & \textbf{Total Duration (hours)} \\
\hline
Xeno-canto$^{*}$ & Yes & 585 & 127,161  & 1802 \\
Xeno-canto$^{*}$ & No & 438 & 120,807 & 1726 \\
Observation.org & No &406 & 65,075 & 385 \\
Dawn Chorus & Yes & 50 & 236 & 4 \\
\hline
\vspace{1em}
\end{tabular}
\caption{Summary of sound files and duration for the datasets used in this work. Datasets marked with an asterisk (*) have all classes derived from XC, except for the noise class. The noise class, representing non-bird sounds, is derived from the \textit{warblrb10k} dataset, specifically from the subset labeled as "0" indicating "no birds present".}
\label{table:dataset_summary}
\end{table}

In the following we report on two studies of neural networks, based on slightly differing datasets.
First, \textit{multi-class} classification, based on the assumption that only one foreground bird is active.
To achieve a minimum requirement of at least 10 recordings per species in order to train the recognition algorithm, we curated a list of 438 bird species for the multi-class model (AvesEcho v0).
Second, to train a \textit{multi-label} model in which multiple species can be simultaneously active, we relaxed the selection criteria to include more species from the EBBA2 list, and also to include recordings from outside Europe for species where we did not have enough recordings. Additionally, we added recordings from outside Europe for all species with less than 50 recordings to make the total for a species up to 50. This curation resulted in a list of 585 species for the multi-label model (AvesEcho v1).

For both of these datasets, a noise class is included, which is derived from the \textit{warblrb10k} dataset\footnote[1]{\url{https://dcase.community/challenge2018/task-bird-audio-detection#audio-datasets}}, specifically from the subset labeled as "0" indicating "no birds present," and so consisting just noise.
\begin{figure}[ht!] 
\centering
\includegraphics[width=3in]{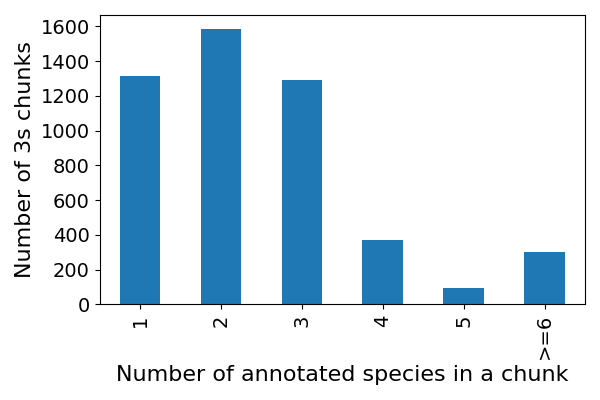}
\caption{Histogram depicting the number of labeled species within 3-second chunks from the Dawn Chorus dataset, illustrating the degree of polyphony present.}
\label{dawnchorus_polyphony}
\end{figure}


We also present results using a Dawn Chorus dataset developed by the Dawn Chorus project\footnote[2]{\url{https://dawn-chorus.org/}}, a Citizen Science and Art initiative by the Naturkundemuseum Bayern/BIOTOPIA LAB and the LBV (The Bavarian Association for the Protection of Birds and Nature). The recordings were mostly short smartphone-based soundscapes collected during the dawn chorus – a time when many birds are most vocally active – by a worldwide community of enthusiasts. The dataset used in this study is a collection of 1-minute bird vocalisation recordings annotated by volunteers, including both professional and lay ornithologists. Only annotations from volunteers who scored 70\% or higher on a species recognition questionnaire were included to ensure a high level of accuracy. The dataset comprises 236 annotated recordings from Germany, selected from a large worldwide pool of recordings\footnote[3]{\url{https://explore.dawn-chorus.org}} made available through the Dawn Chorus project. Annotations are conducted in 3-second chunks, following the BirdNET scheme, and are based on species identification through audio cues. The recordings in the dataset are primarily in uncompressed WAV format, although other formats like MP3 and FLAC are also present, particularly from earlier collection efforts. This subset represents a diverse range of bird species from Germany. Each recording in the dataset is linked to additional metadata, such as location and environmental conditions, though these were not utilized in the current analysis.

\subsubsection{Challenges with Xeno-canto data for model training}
While Xeno-canto offers significant advantages, such as being the largest open-access database of nature sounds, covering all European breeding birds, and providing high-quality, peer-reviewed, and reliable identifications, there are several challenges when using this data for training machine learing models.
\begin{figure}[ht!] 
\centering

\begin{subfigure}[b]{0.49\textwidth}
    \centering
    \includegraphics[width=\textwidth]{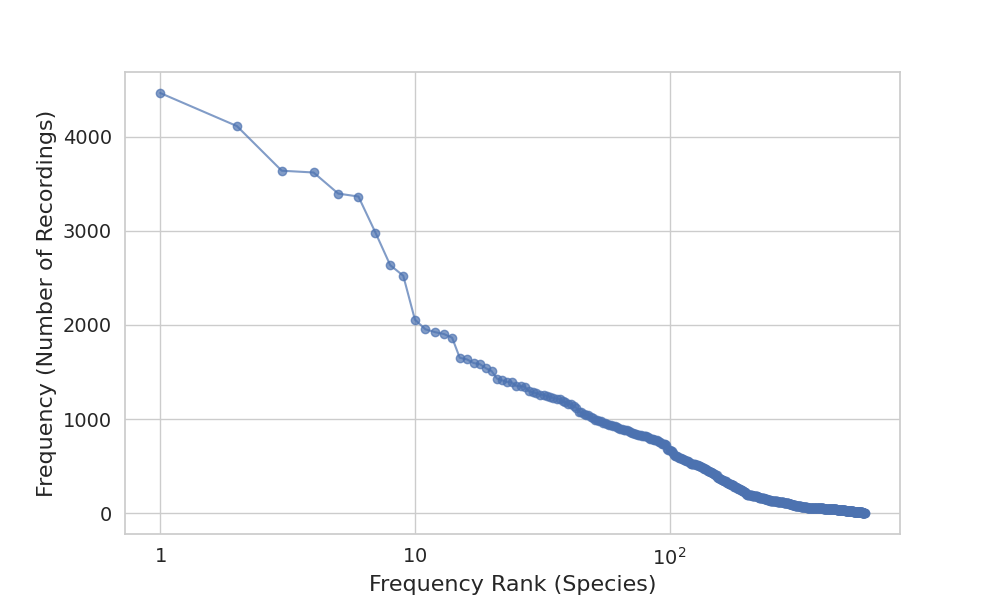}
\end{subfigure}
\hfill
\begin{subfigure}[b]{0.49\textwidth}
    \centering
    \includegraphics[width=\textwidth]{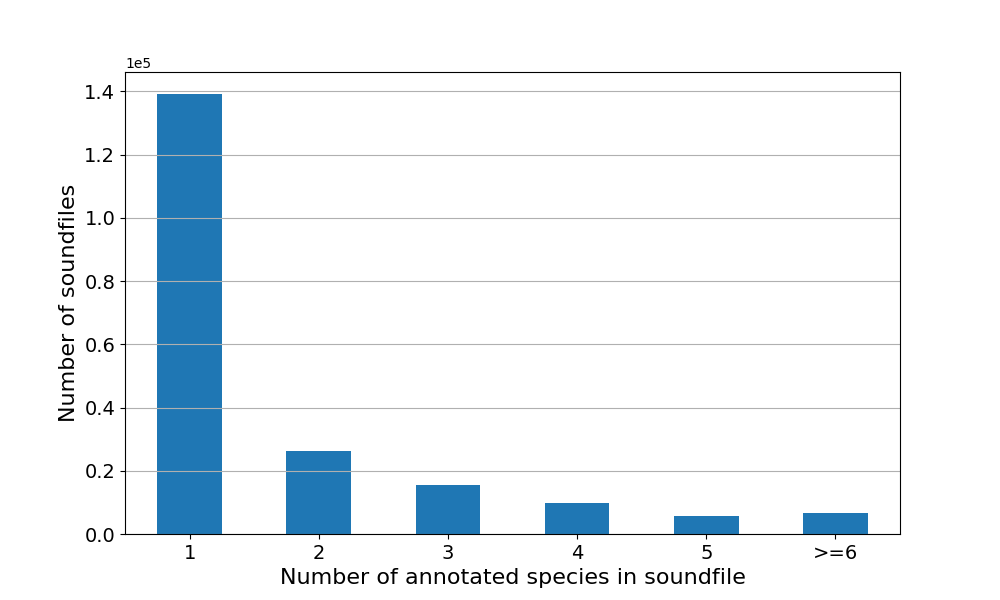}
\end{subfigure}

\caption{(Left) Distribution of the number of recordings per species in the Xeno-Canto dataset for multi-label case. The y-axis represents the count of recordings per species. The mark within the box indicates the median value, and the box itself represents the interquartile range (middle 50\% of the species). (Right) Distribution of the total number of species annotated per file in our Xeno-canto data. The large peak at 1 indicates that many sound files are annotated as having no other species audible in the background.}
\label{overall_distribution}
\end{figure}

Firstly, the data is weakly labelled: while the species of the vocalizing bird is known, the exact timing of these vocalizations within the recordings is not, complicating segment-based model training.
Secondly, many background species are missing in the annotations, potentially impairing the detection of background vocalizations. This arises from Xeno-canto's history, with its original design allowing only for one species per file. For machine learning this is a type of \textit{label noise}: since many occurrences are in effect mislabelled as absences, it can bias a learning algorithm against making correct positive predictions.
Thirdly, Xeno-Canto recordings are primarily focal recordings, designed to capture a target bird's sound as clearly as possible, making these vocalizations louder and more prominent than those typically found in soundscapes. This discrepancy poses difficulties in recognizing fainter bird sounds.
Finally, the dataset is limited for certain species, such as \textit{Stercorarius pomarinus} and \textit{Puffinus mauretanicus}, which have very few recordings available. Fig.~\ref{overall_distribution} illustrates the distribution of the number of recordings per species within the dataset curated to train the model. This distribution highlights the variability in the amount of available data across different bird species. This scarcity necessitates either the removal of these classes or the collection of additional data to ensure accurate classification. 

Despite these limitations, Xeno-canto powers most general-purpose birdsong classifiers, and this yields better bioacoustic recognition than models trained on more generic audio data such as AudioSet~\cite{ghani2023global}.
In the present study we aimed to evaluate how to make best use of the characteristics of Xeno-canto data, including the possible use of background labels.

\subsection{Pretrained models}
\textbf{BirdNET}~\cite{kahl2021birdnet} utilizes EfficientNet, a well-established CNN architecture. It has been trained on an extensive dataset that includes recordings from Xeno-canto (XC), the Macaulay Library, and labeled soundscape data from various global locations, enabling it to recognize many thousands of bird species. Beyond avian identification, BirdNET is also capable of identifying human speech, dogs, and numerous frog species. This broad training set allows BirdNET to be highly versatile, supporting a range of downstream applications. To achieve this versatility, BirdNET strikes a balance between accuracy and computational efficiency, ensuring it can be deployed on a variety of devices with limited processing power. In this work, we report on BirdNET version 2.2, which offers improved performance and broader species coverage. The BirdNET codebase, along with comprehensive documentation, is publicly available on GitHub\footnote{\url{https://github.com/kahst/BirdNET-Analyzer/}}, facilitating further research and development in bioacoustic monitoring and related fields.\\
\textbf{PaSST}~\cite{koutini2021efficient} leverages the strengths of transformer architectures for audio classification. The model, similar to the AST~\cite{gong2021ast}, is inspired by the Vision Transformer (ViT)~\cite{dosovitskiy2020image}. It extracts patches from the spectrogram, adds trainable positional encodings, and applies transformer layers to the flattened sequence of patches. Transformer layers utilize a global attention mechanism, which causes the computation and memory requirements to increase quadratically as the input sequence lengthens~\cite{schmid2023efficient}. PaSST attempts to mitigate this by introducing an innovative training technique known as "Patchout," which selectively drops patches of the spectrogram during training, enhancing efficiency and robustness. In this way, the model significantly reduces compute complexity and memory requirements for training these transformers for audio processing tasks. The model is competitive with SOTA and obtains a mAP of 47.1 using their best single model on the AudioSet-2M. The implementation of PaSST, along with its training framework and performance benchmarks is available for public access on GitHub\footnote{\url{https://github.com/ kkoutini/PaSST}}. \\
\textbf{PSLA}~\cite{gong2021psla} is an EfficientNet model trained on AudioSet. We include this model to have a comparison with the BirdNET model which employs the same CNN-based architecture. By pretraining on large-scale audio datasets, PSLA captures a wide range of audio features, which are further refined through strategic sampling and labeling processes. To generate predictions, it uses an attention layer on the final embeddings, distinguishing it slightly from BirdNET, which uses a single dense layer for the same purpose. The model is again competitive with SOTA and obtains a mAP of 44.3 using their best single model on the AudioSet-2M. In this work, PSLA is incorporated into our pipeline as simply a model pre-trained on AudioSet, and we apply fine-tuning after our own pre-processing steps, rather than using the pre-processing pipeline followed in the PSLA paper. The PSLA framework, along with its implementation details and performance metrics is documented and accessible for public access on GitHub\footnote{\url{https://github.com/YuanGongND/psla}}.

\subsection{Data preprocessing}

For our experiments, we used two different configurations for generating mel-spectrograms, tailored to the requirements of the PaSST and PSLA models. For the PaSST configuration, we used 128 mel bands, a sampling rate of 16,000 Hz, a window length of 400 samples, a hop size of 160 samples, and an FFT size of 512. With a 3-second input signal, these settings result in a mel-spectrogram with 298 time frames. For the PSLA configuration, we used 128 mel bands, a sampling rate of 32,000 Hz, a window length of 800 samples, a hop size of 320 samples, and an FFT size of 1024. Similarly, a 3-second input signal with these settings also results in a mel-spectrogram with 298 time frames. Both configurations ensure that the mel-spectrograms have the same number of time frames, facilitating a consistent comparison between the models. 
Unlike PaSST and PSLA, BirdNET includes a pre-processing stem (a spectrogram layer). For BirdNET, the data is fed as an audio time series with a fixed input size of 3 seconds, using a sampling rate of 48,000 Hz. 

Additionally, we also experimented with various data augmentation techniques to improve model robustness and performance. After evaluating several options, we found that two augmentations provided the best results: MixUp~\cite{zhang2017mixup} and background noise addition. Both of these augmentations are applied directly to the raw waveforms before spectrogram conversion.

MixUp is a data augmentation technique where two samples are combined to create a new synthetic sample. We apply MixUp with a probability of 0.6, meaning that 60\% of the time, two random samples are mixed together to create a new training example. A value \textit{lam} is sampled from a Beta distribution to linearly combine two input samples and their labels, ensuring smooth interpolation between them rather than a hard selection, with \textit{lam} controlling the ratio of the mix.

For background noise addition, we augment the input sample by mixing them with noise samples with a probability of 0.5. In this way, a random half of training samples are mixed with noise. The noise samples are derived from the \textit{warblrb10k} dataset, which contains a variety of background sounds, ensuring that the model becomes more resilient to environmental noise and can better generalize to real-world conditions.

The probability values for both MixUp and background noise were manually tuned in pilot testing. By incorporating these augmentations, we aim to enhance the models' ability to generalize, ultimately leading to more robust performance across different acoustic environments.

\section{Experimental setup}

When training a classifier, one must choose whether to train for single-label prediction, with the assumption that only one label is active, or for multi-label prediction. Single-label is appropriate for focal recordings in which there is one target of interest, which is the case for a large fraction of Xeno-canto recordings (Fig \ref{overall_distribution}), but multi-label is appropriate for soundscapes more generally, as multiple species can be simultaneously active. We investigated both configurations, using single-label as a first study and focusing on multi-label for our more detailed study.

For evaluation, we process the recordings in 3-second chunks and apply max pooling across all chunks within a recording. Instead of evaluating detections on each chunk individually, we aggregate the detections across the entire recording. This approach is meaningful given that our test set from Xeno-Canto is weakly labeled. By max pooling, we can confidently say that if a species is detected in any of the chunks, it is present in the recording as a whole. Secondly, by using max pooling, we are not merely aggregating detections from all chunks, rather, we are specifically reporting the highest detection score for each species across all chunks of the recording. This approach gives us a more robust and confident measure of the species' presence, as it focuses on the strongest evidence within the entire recording. 

For single-label classification, this means we output one label with the highest score across the chunks, while for multi-label classification, we capture multiple labels, each with their highest scores across all chunks. 

\subsection{Single-Label classification}

To establish a clear starting point and maintain simplicity, we employed single-label classification (multi-class), assuming only one species is active at any given time. This assumption aligns with the majority of items in our training data. We explored various transfer learning paradigms, transferring from BirdNET, which is trained exclusively on bird sounds, as well as models trained on AudioSet, a dataset of general audio.

Each model (both CNN and transformer-based) takes a Mel-spectrogram as an input. Among the transfer learning techniques, we experimented with shallow fine-tuning (where only the last layer is trained while the pre-trained backbone remains frozen), deep fine-tuning (where all parts of the model are trained), and teacher-student model distillation (where training is transferred from a pre-trained model, the teacher, to another model, the student). 

We have used the F1-score to report performance for single-label classification, as it provides a straightforward and balanced measure of precision and recall. A model's prediction for a given audio clip is selected as simply the one species that is most strongly predicted.

\subsection{Multi-Label classification}

In many soundscapes, multiple bird species vocalize simultaneously. Consequently, unlike the previous single-label model, we trained our classifiers in a multi-label fashion. The training data typically include either a single species or one "foreground" species along with some labeled background species, prompting us to explore the best ways to utilize such annotations. Additionally, we expanded the number of classes for this model to 585.
Pretraining is commonly used to enhance classifier performance. For birdsong audio, the question arises whether pretraining is necessary, and if so, whether it should be derived from birdsong or more general audio. Ghani et al. (2023)~\cite{ghani2023global} demonstrated that pretrained features from BirdNET are both powerful and flexible. Therefore, we compared BirdNET pretraining (birdsong-based) with AudioSet pretraining (based on general audio from human soundscapes).
These considerations and combinations were explored by training multiple variants of the machine learning algorithm, using the same standard training set and protocol.

Evaluating multi-label classification is slightly more involved than single-label. Evaluation can be based on thresholded predictions (i.e.\ the predicted species are all species for which the model output exceeds a fixed threshold value), or on the continuous-valued model outputs themselves.
Thresholded metrics (such as accuracy, precision, and recall) lead to results strongly dependent on the chosen decision threshold. Selecting an optimal threshold is complex and varies based on the application, such as prioritizing recall for detecting rare species or precision for monitoring diverse species.
To avoid the challenges of threshold selection, for our main evaluation we favour threshold-free metrics: the area under the ROC curve (AUROC) and average precision (AP; mAP when averaging the AP for all classes). These provide a more general evaluation by integrating performance across all possible thresholds, making them ideal for robust model comparison~\cite{van2024birds}. Besides, AUROC, also has a helpful probabilistic interpretation: it represents the likelihood that a randomly selected positive example is ranked above a negative one~\cite{hanley1982meaning}. For these reasons, we have chosen to use AUROC and mAP for our main evaluation, though we also use thresholded metrics in some targeted investigations of differences between models.

\subsection{Effect of secondary labels}
We further designed some conditions to explore the impact of training with secondary (``background'') labels from the Xeno-canto database. Given that many background species are missing in the annotations, this creates label noise that could bias the learning algorithm. We aimed to determine if incorporating these secondary labels can enhance the detection of background vocalizations, despite their potential for noise. Our experiments compared models trained with primary labels only against those trained with both primary and secondary labels to assess the overall effect on model performance.

Additional to using secondary labels in the \textit{training} phase, we also investigated how our a trained model performs on the \textit{test} set, contrasting the test set items where the target species is labelled as primary against those where it is labelled as secondary. For this analysis, we selected recordings from the test set where only two species were labeled: one as the primary (foreground; main focus of the recording) and the other as the secondary (background species). For each species in our selected subset, we examined the model's recall in two scenarios:
\begin{itemize}
    \item Background Instances: Where the species occurs as a single background species in the recording in the test set as per the annotations.
    \item Foreground Instances: Where the species is the main focus of the recording, with just one other species labeled in the background.
\end{itemize}
This selection allowed us to constrain the experiment, ensuring that we could directly assess the impact of using secondary labels on the model's performance for both foreground and background instances of the same species.
Any differences in performance would then indicate differences in acoustic characteristics, for the same bird labelled in differing situation.

\subsection{Weak labelling: effect of segment selection}

Xeno-canto is a weakly labeled database, where the species of the vocalizing bird is known, but the exact timing of these vocalizations within the recordings is not. This complicates segment-based model training. In this section, we examine the impact of using a detector to discard segments that are likely devoid of bird sounds. By filtering out these no-bird segments, we aim to assess whether this preprocessing step improves the training performance of our models. Our experiments compared model performance with and without the use of the detector to determine its effectiveness in enhancing detection performance.

To extract 3-second segments from our Xeno-Canto dataset using a detector, we employed two methods: energy-based and ML-based.

The energy-based method is adapted from a previously established approach~\cite{boudiaf2023search}. This method uses a energy-based heuristic to identify and select segments likely to contain bird vocalizations while discarding those with non-bird sounds. For further details on the extraction process, please refer to Appendix B in~\cite{boudiaf2023search}.

In the ML-based method we employed a pre-trained deep learning model PANNs~\cite{kong2020panns} which is trained using AudioSet. We selected the chunks for which the score for the \textit{bird} class crossed 0.3.

\section{Results}
\subsection{Single-Label classification}

We trained four different single-label classifiers using the set of transfer learning techniques, then evaluated them using held-out Xeno-canto data (Table~\ref{table:singlelabel_results}).
%
%
\begin{table}[h!]
\centering
\rowcolors{0}{}{} 
\begin{tabular}{>{\raggedright\arraybackslash}p{1.5cm} >{\raggedright\arraybackslash}p{1.5cm} >{\raggedright\arraybackslash}p{3cm} c c}
\hline
\textbf{Model} & \textbf{Pretraining source} & \textbf{Transfer method} & \multicolumn{2}{c}{\textbf{F1-score}} \\
 &  &  & \textbf{Xeno-canto} & \textbf{Observation.org} \\
\hline
PaSST & BirdNET & Knowledge distillation & \textbf{0.704} & 0.49 \\
PaSST & AudioSet & Deep finetuning & 0.679 & 0.45 \\
BirdNET & BirdNET & Shallow finetuning & 0.676 & \textbf{0.52} \\
PSLA & AudioSet & Deep finetuning & 0.632 & 0.42 \\
\hline
\vspace{1em}
\end{tabular}
\caption{Performance of models on held-out test dataset from Xeno-canto and the dataset derived from Observation.org for single-label classification. The table compares F1-score across different models and transfer learning methods. Entries are bold-faced if the model scored the highest for the respective dataset.}
\label{table:singlelabel_results}
\end{table}
The best-performing model in our experiments was the PaSST, where knowledge was distilled from BirdNET, achieving an F1-score of 0.704. This suggests that leveraging the domain-specific knowledge embedded in BirdNET through distillation can effectively enhance the performance of a transformer-based architecture like PaSST. The second model involved deep finetuning of all layers of the PaSST model pretrained on AudioSet, which yielded an F1-score of 0.679. Although this model did not outperform the knowledge-distilled PaSST, it still demonstrated robust performance. The third model, which involved training a dense layer on top of the BirdNET backbone, resulted in an F1-score of 0.676. Notably, this score is nearly identical to that of the PaSST model pretrained on AudioSet and deeply finetuned. This suggests that BirdNET, with its architecture and pretraining specifically on bird vocalizations, remains highly competitive even with minimal additional training (shallow finetuning). The fourth model, a PSLA architecture pretrained on AudioSet and deeply finetuned, achieved an F1-score of 0.632. This was the lowest score among the models tested, indicating a significant drop in performance compared to the other models. 

The performance of our models was also evaluated on a separate dataset extracted from \textit{Observation.org}, which, similar to Xeno-Canto, consists of focal recordings. This allowed us to test the models in a multi-class scenario. As shown in Table~\ref{table:singlelabel_results}, the highest F1-score was achieved using the shallow finetuned model where a dense layer was trained on top of the BirdNET backbone, scoring 0.52. The knowledge-distilled transformer model (PaSST with BirdNET pretraining) performed second best with an F1-score of 0.49, followed by the deeply finetuned PaSST model pretrained on AudioSet with a score of 0.45. The PSLA model, also deeply finetuned on AudioSet, had the lowest performance, with an F1-score of 0.42. These results further emphasize the effectiveness of shallow finetuning on bird-specific pretraining, particularly for multi-class bird species classification tasks.  

\begin{figure}[h!]
\centering
\fbox{ 
\begin{minipage}{0.97\textwidth} 
    \centering
    \begin{tabular}{>{\raggedright\arraybackslash}p{1.3cm} >{\raggedright\arraybackslash}p{1.5cm} >{\raggedright\arraybackslash}p{3cm} c c c >{\raggedright\arraybackslash}p{1.2cm} >{\raggedright\arraybackslash}p{1cm}}
    \hline
    \textbf{Model} & \textbf{Pretraining source} & \textbf{Transfer method} & \textbf{mAP} & \textbf{AUROC} & \textbf{Epochs} & \textbf{Time/epoch (s)} & \textbf{Total train time (hours)} \\
    \hline
    PaSST & BirdNET & Knowledge distillation & \textbf{0.71} & \textbf{0.95} & 10 & 8386 & 23.29 \\
    PSLA & BirdNET & Knowledge distillation & 0.67 & 0.94 & 8 & 2307 & 5.21 \\
    BirdNET & BirdNET & Shallow finetuning & 0.68 & 0.95 & 2 & 1060 & 0.59 \\
    PaSST & AudioSet & Shallow finetuning & 0.43 & 0.91 & 30 & 3449 & 28.74 \\
    PSLA & AudioSet & Shallow finetuning & 0.03 & 0.71 & 10 & 1040 & 2.88 \\
    PaSST & AudioSet & Deep finetuning & 0.64 & 0.92 & 8 & 8138 & 18.09 \\
    PSLA & AudioSet & Deep finetuning & 0.64 & 0.94 & 21 & 2188 & 12.76 \\
    \hline
    \end{tabular}

    \vspace{1em}

    \includegraphics[width=5.3in]{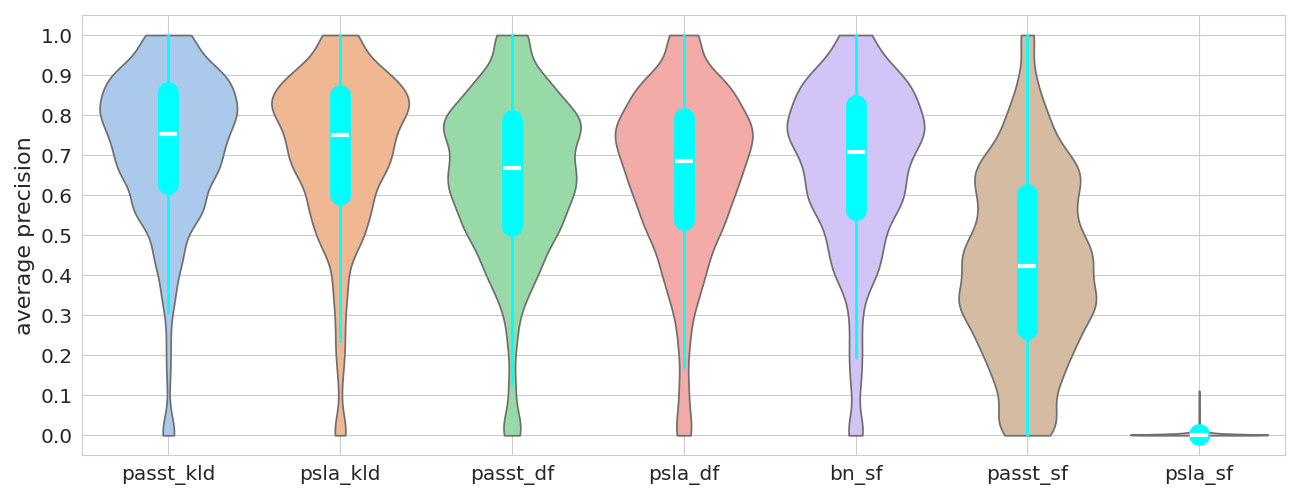}
\end{minipage}
} 
\caption{Performance of different models evaluated on the held-out test dataset from Xeno-canto. (a) Table showing various metrics including mAP, AUROC, epochs, time per epoch, and total training time. (b) Species-wise distribution of scores using the approaches listed in (a). Entries are bold-faced if the model scored the highest.}
\label{fig:table_and_violin}
\end{figure}
\subsection{Multi-Label classification}
We trained multi-label classifiers using the set of transfer learning techniques, again evaluating them on a held-out Xeno-canto test set
(Table in Fig.~\ref{fig:table_and_violin}).

%
Similarly as the single-label case,
the PaSST model with knowledge distillation from BirdNET achieved the best performance, with a mAP of 0.71 and an AUROC of 0.95, requiring 10 epochs and just over 23 hours of total training time. The BirdNET model with shallow finetuning, while having a slightly lower mAP (0.68), matched the AUROC (0.95) and completed training significantly faster, in less than 1 hour. Other models, such as the PSLA and PaSST models pretrained on AudioSet, showed varying performance, with the PSLA model using shallow finetuning on AudioSet scoring the lowest with a mAP of 0.03. Overall, the results highlight the effectiveness of knowledge distillation from BirdNET, particularly with the PaSST model, in achieving high multi-label classification performance. The inability of the shallow fine-tuned PSLA to train might suggest that the embeddings' space in PSLA is heavily influenced by the attention pooling layer, perhaps making it intrinsically tied to the specific dataset used during training. 

Inspecting the species-wise distribution of scores using different approaches, we found that the distribution of performance per-species was generally consistent between models, and coherent with the overall performance scores (Figure~\ref{fig:table_and_violin}).


\begin{table}[h!]
\centering
\rowcolors{0}{}{} 
\begin{tabular}{>{\raggedright\arraybackslash}p{1.5cm} >{\raggedright\arraybackslash}p{3cm} >{\raggedright\arraybackslash}p{3cm} >{\raggedright\arraybackslash}p{2cm} >{\raggedright\arraybackslash}p{2cm}}
\hline
\textbf{Model} & \textbf{Pretraining source} & \textbf{Transfer method} & \textbf{mAP} & \textbf{AUROC} \\
\hline
BirdNET & BirdNET & Shallow finetuning  & \textbf{0.311} & \textbf{0.836} \\
PaSST & BirdNET & Knowledge distillation & 0.293 & 0.792 \\
PSLA & BirdNET & Knowledge distillation & 0.273 & 0.793 \\
PaSST & AudioSet & Deep finetuning & 0.291 & 0.794 \\
PSLA & AudioSet & Deep finetuning & 0.256 & 0.768 \\
\hline
\vspace{1em}
\end{tabular}
\caption{Comparison of models on the Dawn Chorus dataset. Entries are bold-faced if the model scored the highest.}
\label{tab:models_comparison_dawnchorus}
\end{table}
We then evaluated our multi-label models on the dawn chorus dataset, to evaluate how well they generalise to new conditions and polyphonic multi-species soundscapes (Table \ref{tab:models_comparison_dawnchorus}).
The BirdNET model with shallow finetuning achieved the highest mAP (0.311) and AUROC (0.836), outperforming the other models. The PaSST model, which had knowledge distilled from BirdNET, performed slightly lower with a mAP of 0.293 and an AUROC of 0.792. Similarly, the PSLA model using knowledge distillation from BirdNET showed comparable results with a mAP of 0.273 and AUROC of 0.793. Models pre-trained on AudioSet, specifically PaSST and PSLA with deep finetuning, scored lower, with the PSLA model obtaining the lowest mAP (0.256) and AUROC (0.768).

\begin{figure}[ht!] 
\centering
\includegraphics[width=4.5in]{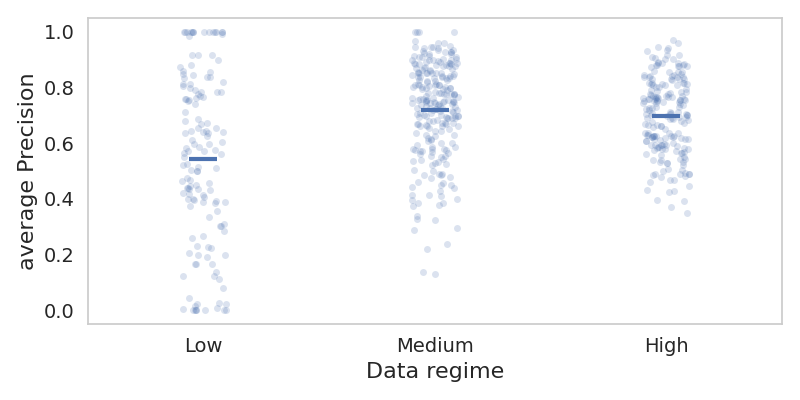}
\caption{Distribution of species-specific average precision scores using the shallow fine-tuned BirdNET model, grouped by data availability for each species (low, medium, and high data regimes based on the amount of available data in the XC-derived dataset).}
\label{data_regimes}
\end{figure}
In order to understand the importance of data quantity in training a robust multi-label model, we performed an analysis of the model's performance across different data regimes (Figure~\ref{data_regimes}). The results reveal a clear dependency on the amount of training data available for each species. Species with a high number of recordings consistently achieved higher average precision scores, indicating strong model performance in this regime. In contrast, species in the low data regime, with as few as 10 recordings, showed significantly lower average precision and greater variability, highlighting the model's difficulty in accurately predicting species with limited data. The medium data regime displayed intermediate results, with moderate average precision and some variability, further underscoring the importance of data quantity in achieving reliable model performance. These findings emphasize the critical role of sufficient training data in enhancing model performance for multi-label bird species classification.

\subsection{Effect of secondary labels}

In this experiment, we explored the impact of incorporating secondary labels while training on the performance of our bird species recognition model. Our aim was to determine whether these additional labels, despite the label noise, could enhance the model's ability to detect background vocalizations. The results of this experiment are summarized in Table \ref{tab:models_comparison}. The overall mAP remained essentially unchanged (0.68 vs. 0.68 for XC, 0.31 vs. 0.31 for dawn chorus), indicating that the introduction of secondary labels did not significantly affect the model's ability to rank predictions accurately across the datasets.
\begin{table}[h!]
\centering
\rowcolors{0}{}{} 
\begin{tabular}{>{\raggedright\arraybackslash}p{3cm} >{\raggedright\arraybackslash}p{2.5cm} >{\raggedright\arraybackslash}p{1.5cm} >{\raggedright\arraybackslash}p{1.5cm} >{\raggedright\arraybackslash}p{1.5cm} >{\raggedright\arraybackslash}p{1.5cm}}
\hline
\textbf{Dataset} & \textbf{Train using secondary labels} & \textbf{Precision} & \textbf{Recall} & \textbf{mAP} & \textbf{AUROC} \\
\hline
\multirow{1}{*}{Xeno-Canto} & Yes & 0.44 & 0.70  & 0.68 & 0.96 \\
 & No & 0.59 & 0.61 & 0.68 & 0.95 \\
\hline
\multirow{1}{*}{Dawn chorus} & Yes & 0.74 & 0.45  & 0.31 & 0.84 \\
& No & 0.88 & 0.14 & 0.31 & 0.81 \\
\hline
\vspace{1em}
\end{tabular}
\caption{Comparison of training with and without using secondary labels on the XC-derived test set and the dawn chorus datasets. The reported performance is using the shallow finetuned BirdNET model. To compute the precision and recall, a common threshold of 0.2 was set.}
\label{tab:models_comparison}
\end{table}

However, we observed a notable decrease in precision for both the XC-derived test set (0.44 with secondary labels vs. 0.59 without) and the dawn chorus soundscapes dataset (0.74 with secondary labels vs. 0.88 without) when using secondary labels. This suggests that the model became more prone to predicting the presence of species even in cases where they were not present, likely due to the noise and incompleteness of these labels. 
\begin{figure}[ht!] 
\centering
\includegraphics[width=5in]{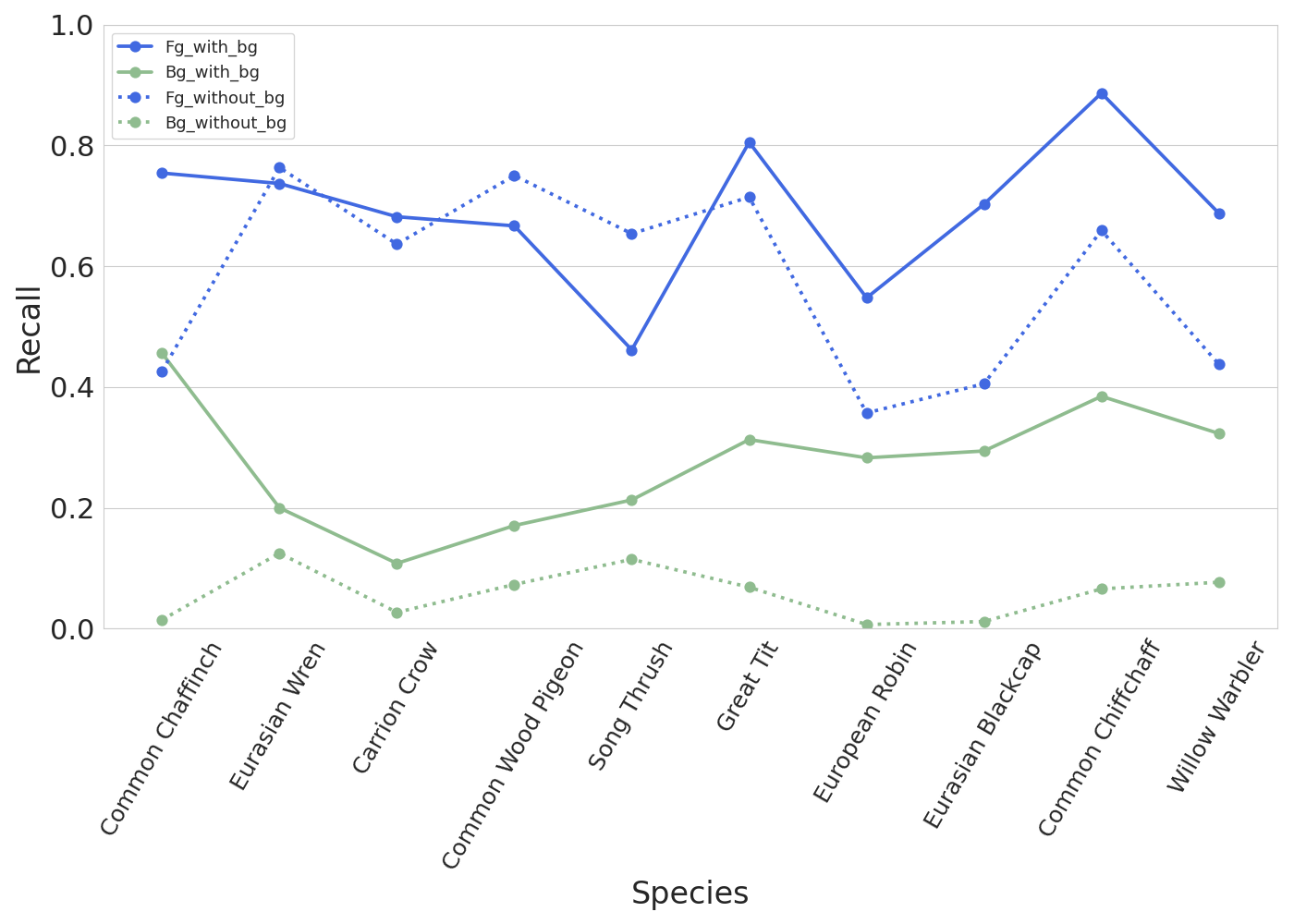}

\caption{Recall of foreground and background predictions for common bird species in our XC-derived test set, where exactly two species are labeled as present: one in the foreground and one in background. For each species, recall is measured when instances are labeled as either foreground (blue) or background (green). Solid lines represent predictions from a model trained using both foreground and background labels, while dotted lines represent predictions from a model trained using only foreground labels. In both cases, the model is a shallow fine-tuned BirdNET.}
\label{forg_backg_comparison_2sp_w_wo_training}
\end{figure}

Conversely, recall increased in both test scenarios  (0.70 with secondary labels vs. 0.61 without for XC, and 0.45 with secondary labels vs. 0.14 without for dawn chorus). This indicates that the model became more sensitive to detecting species, successfully identifying more true positives, including those that may have been underrepresented or missing from the primary labels. 

The AUROC score, on the other hand, increased when incorporating secondary labels (0.96 with secondary labels vs. 0.95 without for XC, and 0.84 with secondary labels vs. 0.81 without for dawn chorus). This suggests an overall improvement in the model's ability to distinguish between species presence and absence across various decision thresholds, indicating enhanced discrimination capability despite the label noise.

To investigate this balance in more detail, we selected 10 common species for analysis, then analysed model predictions for all Xeno-canto sound recordings that contain one of these 10 species as primary or secondary, restricting our attention to the sound recordings containing exactly two species in total.
This scenario enables us to probe how predictive performance varies whether a sound is primary or secondary (implying that it may be foreground or background).
For all 10 species, the model's accuracy in identifying instances where the species was labeled as secondary improved when the model was trained using secondary labels (Fig.~\ref{forg_backg_comparison_2sp_w_wo_training}). This suggests that the inclusion of secondary labels during training helped the model better recognize species even when they are not the primary focus of a recording. The results for foreground instances were more mixed. For some species, the accuracy increased when secondary labels were included in the training phase, indicating that the model became better at recognizing these species even when they were in focus. However, for other species, the accuracy of foreground instances decreased. This mixed result suggests that, while the model's overall sensitivity improved, the specific characteristics of certain species when they are the focus of a recording might have been less distinguishable when trained with noisy or incomplete secondary labels.


\subsection{Weak labelling: effect of segment selection}
We compared the performance of models trained with and without the use of the detector, evaluating their performance on both the XC-derived test set and the dawn chorus datasets. The results, as shown in Table~\ref{tab:models_comparison_weak} indicate that the effect of segment selection using a detector is at best marginal. For the XC-derived test set, training with the energy-based detector yielded a slight improvement in mAP (0.70 vs. 0.68), while the AUROC remained unchanged at 0.95. Similarly, for the dawn chorus dataset, using the energy-based detector led to a small increase in both mAP (0.32 vs. 0.31) and AUROC (0.82 vs. 0.81). Surprisingly, we did not notice any noticeable gain when employing the PANNs detector for selecting chunks. 
\begin{table}[h!]
\centering
\rowcolors{0}{}{} 
\begin{tabular}{>{\raggedright\arraybackslash}p{3cm} >{\raggedright\arraybackslash}p{2.5cm} >{\raggedright\arraybackslash}p{1.5cm} >{\raggedright\arraybackslash}p{1.5cm} >{\raggedright\arraybackslash}p{1.5cm} >{\raggedright\arraybackslash}p{1.5cm}}
\hline
\textbf{Dataset} & \textbf{Detection} & \textbf{mAP} & \textbf{AUROC} \\
\hline
\multirow{3}{*}{Xeno-Canto} & Peak detector & 0.70 & 0.95 \\
 & PANNs detector & 0.67 & 0.95 \\
 & No & 0.68 & 0.95 \\
\hline
\multirow{3}{*}{Dawn chorus} & Peak detector & 0.32 & 0.82 \\
 & PANNs detector & 0.30 & 0.82 \\
 & No & 0.31 & 0.81 \\
\hline
\vspace{1em}
\end{tabular}
\caption{Comparison of training with and without using detection algorithms on the XC-derived test set and the dawn chorus datasets. The reported performance is using the shallow finetuned BirdNET model.}
\label{tab:models_comparison_weak}
\end{table}

These results suggest that while a detector does contribute to a slight improvement in model performance, the overall effect is small to negligible. This could indicate that the models are robust enough to handle the weak labeling inherent in the Xeno-Canto dataset. Thus, while segment selection might offer a modest boost in performance, its impact may be limited depending on the nature of the dataset and the model used.

\section{Discussion}

\subsection{Transfer learning}
We have shown that transfer learning leads to a consistent strong performance in birdsong classification, across a variety of conditions: both single- and multi-label, both CNN and Transformer models.
This is the case for the common fine-tuning method, but also for knowledge distillation.
In particular, we investigated cross-model distillation, which is knowledge distillation from one type of deep learning architecture to a different one.
We find that cross-model distillation can yield classifiers that give better performance than those trained with fine-tuning, when evaluated in-domain on Xeno-canto data (Table \ref{table:singlelabel_results}, Fig. \ref{fig:table_and_violin}).

However, contrary to indications from related literature on cross-distillation~\cite{gong2022cmkd}, we find that the generalisation to novel soundscape datasets, while strong, does not outperform shallow fine-tuning (Table \ref{table:singlelabel_results}, Table \ref{tab:models_comparison_dawnchorus}). It thus appears that shallow fine-tuning has better generalisation capabilities than knowledge distillation. This is likely caused by the fact that shallow fine-tuning is the more constrained form of training, having fewer degrees of freedom.

Despite this concern, knowledge distillation brings the significant advantage that it liberates us from the design decisions of previous neural network architectures, enabling automatic bioacoustic recognition to keep pace with developments in the general field of machine learning. For example, future researchers will need to consider not only CNNs and Transformers, but also alternatives such as state-space models (Mamba) or graph neural networks~\cite{bhati2024dass, castro2024graph}.

As we have demonstrated, it is feasible to train a bird classifier made from a high-performing modern Transformer architecture (PaSST) while reusing the knowledge contained in a completely different CNN-type architecture (BirdNET).
Knowledge distillation takes more computational effort than shallow fine-tuning---as expected, since the latter trains only a small part of a network---yet this is not the dominant factor for computational load in our studies. Rather, the choice of neural network architecture has the strongest influence on the total training time required.
Thus, future work should investigate how to make use of knowledge distillation for flexible knowledge transfer, while also constraining the algorithm to enhance generalisation.

Our analysis further reveals that while bird-specific pretraining generally leads to superior performance, models pretrained on broader audio datasets like AudioSet also achieve good results. For example, the BirdNET model with shallow finetuning consistently outperforms others, particularly in challenging environments such as the dawn chorus, which is known for its high polyphony. Despite this, models like PaSST pretrained on AudioSet demonstrate that general audio pretraining can still produce competitive results, especially in scenarios where specialized bird sound datasets may not be readily available. This suggests that transfer learning is a flexible method, in which even general audio models can adapt effectively to specific bioacoustic tasks.

While transfer learning is advantageous for developing models in low data regimes, our results reveal that the quantity of data still plays a role in model performance. As shown in Figure~\ref{data_regimes}, species with ample training data achieve strong average precision scores, whereas those with limited data show greater variability—some perform well, but others poorly. This suggests that while transfer learning helps mitigate data scarcity, other factors, such as vocalization complexity, recording quality, background noise, and species-specific variability, might also impact performance. These findings highlight the need to consider data quality and species-specific characteristics alongside transfer learning to achieve reliable model performance, particularly in low-data scenarios.

\subsection{Use of single-species and polyphonic training data}

Xeno-canto is the most important data source for training birdsong classifiers, yet it contains a mixture of single-species and multi-species recordings, the latter often incompletely labelled.
Thus we investigated how best to use this data for recognition in polyphonic birdsong soundscapes.
Our results suggest that while the introduction of secondary labels increases the model's sensitivity (as seen in higher recall), it also introduces a higher risk of false positives (reflected in lower precision). The stability of mAP implies that, although the model's confidence may have decreased, its overall ranking of species likelihoods across the dataset did not suffer. This trade-off between precision and recall must be carefully considered when designing bird species recognition models, especially in scenarios where the completeness and accuracy of labels are in question. 

Model predictions on Xeno-canto data are more confident and more accurate for species that have been labelled as primary (foreground, or focal) species, versus secondary species (Figure \ref{forg_backg_comparison_2sp_w_wo_training}).
This tells us that there is some difference in the acoustic data between these two cases. So what is this difference?
Although the (meta)data do not answer this question explicitly,
community practice in Xeno-canto is to label as primary the most salient species heard, or the species that the recordist was targeting.
Secondary species might be overlapping in time with the primary, which can lead to acoustic masking for a listener, or might be separate in time; the data doubtlessly contain examples of each case.

The more general differences are likely to be the interrelated factors of loudness and distance. Loudness can be probed via the relative amplitude of recordings, though we remind the reader that the data in question are not provided with calibrated amplitudes. Distance affects loudness as well as other characteristics such as reverberation and the atmospheric absorption of high frequencies.
The task of species classification is, formally, not dependent on loudness: sounds should be labelled as present irrespective of their (relative) amplitudes, and machine learning algorithms should then learn to perform the task invariant to amplitudes. However, even with deep learning systems the strength or confidence in detecting a signal is related to the signal-to-noise ratio and the distinctiveness of the signal---though, as for human listeners, this relationship is nonlinear. Further, quieter sounds have a lower probability of having been positively labelled in the training data. Thus in many audio recognition tasks it is typical that the louder a sound, the more confidently an automatic classifier can pick up on it.

To produce discrete predictions, classifier outputs are thresholded in practice. Calibrating thresholds to local data conditions has been identified to be crucial, if automatic recognition is to produce reliable biodiversity observations \cite{Sethi:2024, wood2024guidelines}. In this work we do not attempt to address this calibration issue, but the issue is nonetheless evident in the fact that our results on generalisation to novel soundscapes have relatively low mAP (evaluated on thresholded detections) even when they have strong AUC scores. It remains an open question how far we can automate or avoid such calibration.

Our study finds positive results when using the labels of secondary (``background'') species in Xeno-canto sound recordings, despite the fact that such labels may be incomplete.
Therefore it would be valuable for the Xeno-canto community, and those who wish to make use of its data, to invest effort into labelling the sound recordings more completely.
Our observations demonstrate the value of labelling all species audible in the background; it would also be useful to annotate temporal details (start time and end time of vocalisations), which would help to overcome the issue of weak labelling in the training of classifiers.


\section*{Acknowledgements}
The Dawn Chorus data were collected through the Dawn Chorus Project \url{https://dawn-chorus.org/} by Naturkundemuseum Bayern/BIOTOPIA LAB and LBV. We extend our gratitude to Lisa Gill for enabling us to access the Dawn Chorus annotation data subset. We thank the many volunteers who contributed sound recordings and annotations to Xeno-canto, to Observation.org, to Warblr, and to the Dawn Chorus project.

\section*{CRediT authorship contribution statement}
\textbf{Burooj Ghani:} Conceptualization, Methodology, Software, Validation, Investigation, Data curation, Writing – original draft, Visualization. \textbf{Vincent J. Kalkman:} Conceptualization, Investigation, Data curation, Supervision, Writing – review \& editing, Project administration.
\textbf{Bob Planqué:} Data curation, Writing – review \& editing. \textbf{Willem-Pier Vellinga:} Data curation, Writing – review \& editing. \textbf{Lisa Gill:} Data curation, Writing – review \& editing. \textbf{Dan Stowell:} Conceptualization, Methodology, Investigation, Writing – original draft, Supervision, Project administration.

\section*{Conflict of interest}
None

\section*{Funding}
Burooj Ghani is supported by EU-funded Horizon projects MAMBO (101060639), GUARDEN (101060693) and TETTRIs ().


\bibliography{sample}




\end{document}